\input harvmac
\input amssym
\input epsf

\newcount\figno
\figno=0 
\def\fig#1#2#3{
\par\begingroup\parindent=0pt\leftskip=1cm\rightskip=1cm\parindent=0pt
\baselineskip=11pt
\global\advance\figno by 1
\midinsert
\epsfxsize=#3
\centerline{\epsfbox{#2}}
\vskip 12pt
{\bf Fig.\ \the\figno: } #1\par
\endinsert\endgroup\par
}
\def\figlabel#1{\xdef#1{\the\figno}}

\noblackbox



\lref\MIZA{
J.~A.~Minahan and K.~Zarembo,
JHEP {\bf 0303}, 013 (2003)
[arXiv:hep-th/0212208].
}

\lref\BeisertYB{ N.~Beisert and M.~Staudacher, 
Nucl.\ Phys.\ B {\bf 670}, 439 (2003)
[arXiv:hep-th/0307042]. 
}  

\lref\tseytlin{
A.~A.~Tseytlin,
arXiv:hep-th/0311139.
}

\lref\BoosZQ{
H.~Boos, M.~Jimbo, T.~Miwa, F.~Smirnov and Y.~Takeyama,
arXiv:hep-th/0405044.
}

\lref\kbi{
V.~E.~Korepin,  N.~M.~Bogoliubov and A.~G.~Izergin, {\it Quantum inverse
scattering method and correlation functions,} Cambridge University
Press (Cambridge) 1993.}

\lref\faddeev{L.~D.~Faddeev,
arXiv:hep-th/9605187.
}

\lref\KitanineTJ{
N.~Kitanine, J.~M.~Maillet, N.~A.~Slavnov and V.~Terras,
J.\ Phys.\ A {\bf 35}, L753 (2002)
[arXiv:hep-th/0210019].
}

\lref\KitanineAA{
N.~Kitanine, J.~M.~Maillet, N.~A.~Slavnov and V.~Terras,
Nucl.\ Phys.\ B {\bf 642}, 433 (2002)
[arXiv:hep-th/0203169].
}

\lref\spinspin{N.~Kitanine, J.~M.~Maillet, N.~A.~Slavnov and V.~Terras,
Nucl.\ Phys.\ B {\bf 641}, 487 (2002)
[arXiv:hep-th/0201045].
}

\lref\kitaninedet{
N.~Kitanine, J.~M.~Maillet and V.~Terras,
Nucl.\ Phys.\ B {\bf 554}, 647 (1999)
[arXiv:math-ph/9807020].

\

N.~A.~Slavnov, 
Theor.~Math.~Phys.~79~(1989)~502.
}

\lref\GohmannAV{
F.~G\"ohmann and V.~E.~Korepin,
J.\ Phys.\ A {\bf 33}, 1199 (2000)
[arXiv:hep-th/9910253].
}
 \lref\bpr{ I.~Bena, J.~Polchinski and R.~Roiban, 
Phys.\ Rev.\ D {\bf 69}, 046002
(2004) [arXiv:hep-th/0305116]. 
}   

\lref\gkp{
S.~S.~Gubser, I.~R.~Klebanov and A.~M.~Polyakov, 
Nucl.\ Phys.\ B {\bf 636},
99 (2002) [arXiv:hep-th/0204051]. 
} 

 \lref\ft{ S.~Frolov and A.~A.~Tseytlin, 
JHEP {\bf 0206}, 007 (2002) [arXiv:hep-th/0204226]. 
}   

 \lref\jn{ N.~Mann and J.~Polchinski, 
arXiv:hep-th/0305230. 
}  

\lref\dolan{ L.~Dolan, C.~R.~Nappi and E.~Witten, 
arXiv:hep-th/0401243. 

\

L.~Dolan, C.~R.~Nappi and E.~Witten, 
JHEP {\bf 0310}, 017 (2003) [arXiv:hep-th/0308089]. 
}  

\lref\okuyama{ K.~Okuyama and L.~S.~Tseng, 
arXiv:hep-th/0404190. 
}  

\lref\BeisertEA{ N.~Beisert, S.~Frolov, M.~Staudacher and
A.~A.~Tseytlin, 
JHEP {\bf 0310}, 037 (2003) [arXiv:hep-th/0308117]. 
} 

\lref\ArutyunovRG{ G.~Arutyunov and M.~Staudacher,
JHEP {\bf 0403}, 004 (2004) [arXiv:hep-th/0310182]. 
}  

\lref\kruc{ M.~Kruczenski, 
arXiv:hep-th/0311203; 
\

M.~Kruczenski, A.~V.~Ryzhov and A.~A.~Tseytlin, 
arXiv:hep-th/0403120. 
}

\lref\SerbanJF{ D.~Serban and M.~Staudacher, 
arXiv:hep-th/0401057. 
}   
\lref\XXZ{ P.~Di Vecchia and A.~Tanzini, 
arXiv:hep-th/0405262. 
}
\lref\BECH{
D.~Berenstein and S.~A.~Cherkis, 
arXiv:hep-th/0405215.
}
\lref\RR{ 
R.~Roiban, 
arXiv:hep-th/0312218. 
}
 
\lref\othera{ 
J.~A.~Minahan, 
arXiv:hep-th/0405243. 
}
\lref\otherb{
I.~Swanson, 
arXiv:hep-th/0405172. 
}
\lref\otherc{
L.~Freyhult, 
arXiv:hep-th/0405167. 
}
\lref\otherd{
A.~Agarwal and S.~G.~Rajeev, 
arXiv:hep-th/0405116. 
}
\lref\othere{
A.~M.~Polyakov, 
arXiv:hep-th/0405106. 
}
\lref\otherf{
M.~Smedback, 
Phil.\ Trans.\ Roy.\ Soc.\ Lond.\ A {\bf 356}, 487 (1998)
[arXiv:hep-th/0405102]. 
}
\lref\otherg{
M.~Lubcke and K.~Zarembo, 
arXiv:hep-th/0405055. 
}
\lref\otherh{
N.~Beisert, V.~Dippel and M.~Staudacher, 
arXiv:hep-th/0405001. 
}
\lref\otheri{
A.~V.~Ryzhov and A.~A.~Tseytlin, 
arXiv:hep-th/0404215. 
}
\lref\otherj{
G.~Ferretti, R.~Heise and K.~Zarembo, 
arXiv:hep-th/0404187. 
}
\lref\otherk{
A.~Mikhailov, 
arXiv:hep-th/0404173. 
}
\lref\otherl{
B.~.~J.~Stefanski and A.~A.~Tseytlin, 
arXiv:hep-th/0404133. 
}
\lref\othero{
A.~V.~Kotikov, L.~N.~Lipatov, A.~I.~Onishchenko and
V.~N.~Velizhanin, 
arXiv:hep-th/0404092. 
}
\lref\otherp{
H.~Dimov and R.~C.~Rashkov, 
arXiv:hep-th/0404012. 
}
\lref\otherq{
S.~Ryang,
JHEP {\bf 0404}, 053 (2004) [arXiv:hep-th/0403180]. 
}
\lref\otherr{
R.~Hernandez and E.~Lopez, 
JHEP {\bf 0404}, 052 (2004)
[arXiv:hep-th/0403139]. 
}
\lref\others{
H.~Dimov and R.~C.~Rashkov, 
arXiv:hep-th/0403121. 
}
\lref\othert{
A.~V.~Belitsky,S.~E.~Derkachov, G.~P.~Korchemsky and 
A.~N.~Manashov, 
arXiv:hep-th/0403085. 
}
\lref\otheru{
G.~Arutyunov and M.~Staudacher,
arXiv:hep-th/0403077. 
}
\lref\otherv{
B.~Chen, X.~J.~Wang and Y.~S.~Wu, 
Phys.\ Lett.\ B {\bf 591}, 170 (2004)
[arXiv:hep-th/0403004]. 
}
\lref\otherw{
V.~A.~Kazakov, A.~Marshakov, J.~A.~Minahan and K.~Zarembo, 
arXiv:hep-th/0402207. 
}
\lref\otherx{
J.~Engquist, 
JHEP {\bf 0404}, 002 (2004)
[arXiv:hep-th/0402092]. 
}
\lref\othery{
A.~Mikhailov,
arXiv:hep-th/0402067. 
}
\lref\otherz{
C.~Kristjansen,
Phys.\ Lett.\ B {\bf 586}, 106 (2004)
[arXiv:hep-th/0402033]. 
}
\lref\otheraa{
M.~Alishahiha, A.~E.~Mosaffa and H.~Yavartanoo,
Nucl.\ Phys.\ B {\bf 686}, 53 (2004)
[arXiv:hep-th/0402007]. 
}
\lref\otherab{
O.~DeWolfe and N.~Mann, 
JHEP {\bf 0404}, 035 (2004)
[arXiv:hep-th/0401041]. 
}
\lref\otherac{
B.~Chen, X.~J.~Wang and Y.~S.~Wu, 
JHEP {\bf 0402}, 029 (2004) [arXiv:hep-th/0401016]. 
}
\lref\otherad{
N.~w.~Kim, 
arXiv:hep-th/0312113. 
}
\lref\otherae{
B.~.~J.~Stefanski, 
JHEP {\bf 0403}, 057 (2004) [arXiv:hep-th/0312091]. 
}
\lref\otheraf{
B.~Jurco, 
arXiv:hep-th/0311252. 
}
\lref\otherag{
A.~V.~Belitsky,
S.~E.~Derkachov, G.~P.~Korchemsky and A.~N.~Manashov, 
arXiv:hep-th/0311104. 
}
\lref\otherah{
X.~J.~Wang and Y.~S.~Wu, 
Nucl.\ Phys.\ B {\bf 683}, 363 (2004)
[arXiv:hep-th/0311073]. 
}
\lref\otherai{
A.~Mikhailov,
JHEP {\bf 0312}, 058 (2003)
[arXiv:hep-th/0311019]. 
}
\lref\otheraj{
G.~Arutyunov, J.~Russo and A.~A.~Tseytlin, 
Phys.\ Rev.\ D {\bf 69}, 086009 (2004)
[arXiv:hep-th/0311004]. 
}
\lref\otherak{
N.~Beisert,
Nucl.\ Phys.\ B {\bf 682}, 487 (2004) [arXiv:hep-th/0310252]. 
}
\lref\otheral{
J.~Engquist, J.~A.~Minahan and K.~Zarembo, 
JHEP {\bf 0311}, 063 (2003)
[arXiv:hep-th/0310188]. 
}
\lref\otheram{
N.~Beisert,
JHEP {\bf 0309}, 062 (2003) [arXiv:hep-th/0308074]. 
}
\lref\otheran{
G.~Arutyunov, S.~Frolov, J.~Russo and A.~A.~Tseytlin,
Nucl.\ Phys.\ B {\bf 671}, 3 (2003) [arXiv:hep-th/0307191]. 
}
\lref\otherap{
N.~Beisert, 
Nucl.\ Phys.\ B {\bf 676}, 3 (2004) [arXiv:hep-th/0307015]. 
}
\lref\otheraq{
N.~Beisert, J.~A.~Minahan, M.~Staudacher and K.~Zarembo, 
JHEP {\bf 0309}, 010 (2003)
[arXiv:hep-th/0306139]. 
}
\lref\otherar{
S.~Frolov and A.~A.~Tseytlin, 
JHEP {\bf 0307}, 016 (2003) [arXiv:hep-th/0306130]. 
}
\lref\otheras{
S.~Frolov and A.~A.~Tseytlin, 
Nucl.\ Phys.\ B {\bf 668}, 77 (2003)
[arXiv:hep-th/0304255]. 
}
\lref\otherlast{
N.~Beisert, C.~Kristjansen and M.~Staudacher, 
Nucl.\ Phys.\ B {\bf 664}, 131 (2003)
[arXiv:hep-th/0303060]. 
}      

\lref\GMRII{
D.~J.~Gross, A.~Mikhailov and R.~Roiban,
JHEP {\bf 0305}, 025 (2003)
[arXiv:hep-th/0208231].
}

\lref\KORnorm{
V.~E.~Korepin,
Commun.\ Math.\ Phys.\  {\bf 86}, 391 (1982).
}

\lref\YONE{
S.~Dobashi and T.~Yoneya,
arXiv:hep-th/0406225.
}

\lref\DFMMR{
E.~D'Hoker, D.~Z.~Freedman, S.~D.~Mathur, A.~Matusis and L.~Rastelli,
arXiv:hep-th/9908160.
}

\lref\DEFP{
E.~D'Hoker, J.~Erdmenger, D.~Z.~Freedman and M.~Perez-Victoria,
Nucl.\ Phys.\ B {\bf 589}, 3 (2000)
[arXiv:hep-th/0003218].
}

\lref\GMIT{
N.~R.~Constable, D.~Z.~Freedman, M.~Headrick, S.~Minwalla, L.~Motl,
A.~Postnikov and W.~Skiba, 
JHEP {\bf 0207}, 017 (2002) [arXiv:hep-th/0205089].
}

\lref\BEOP{
N.~Beisert,
Nucl.\ Phys.\ B {\bf 659}, 79 (2003)
[arXiv:hep-th/0211032].
}

\lref\DHoker{
E.~D'Hoker, D.~Z.~Freedman and W.~Skiba,
Phys.\ Rev.\ D {\bf 59}, 045008 (1999)
[arXiv:hep-th/9807098].
}

\Title{\vbox{\baselineskip12pt
	\hbox{hep-th/0407140}
}}{Yang-Mills Correlation Functions from Integrable Spin Chains}

\centerline{
Radu Roiban${}^\dagger$ and
Anastasia Volovich${}^\ddagger$ 
}

\bigskip
\bigskip

\centerline{${}^\dagger$Department of Physics, University of California}
\centerline{Santa Barbara, CA 93106 USA}

\smallskip

\centerline{${}^\ddagger$Kavli Institute for Theoretical Physics}
\centerline{Santa Barbara, CA 93106 USA}

\bigskip
\bigskip

\centerline{\bf Abstract}

\bigskip

The relation between the dilatation operator of ${\cal{N}} = 4$
Yang-Mills theory and integrable spin chains makes it possible to
compute the one-loop anomalous dimensions of all operators in the theory.
In this paper we show how to apply the technology of integrable 
spin chains to the calculation of Yang-Mills correlation functions 
by expressing
them in terms of matrix elements of spin operators on the corresponding
spin chain.  We illustrate this method with several examples in the
$SU(2)$ sector described by the XXX${}_{1/2}$ chain.

\Date{July 2004}

\listtoc
\writetoc

\newsec{Introduction}

Recently there has been considerable progress in studying the AdS/CFT
correspondence beyond the BPS and near-BPS limits (see \tseytlin\ for a
review). Integrability on both sides of the correspondence plays
an important role in these quantitative checks. On the gauge theory
side it has been shown that the planar one-loop dilatation operator
can be identified with the Hamiltonian of an integrable spin chain
\MIZA. The vector space on each site of the spin chain depends on the
particular sector one is interested in. For instance for scalar
operators the dilatation operator can be described by the
$SO(6)$  spin chain in the vector representation \MIZA, while an
$SU(2,2|4)$ spin chain in the singleton representation describes the
full ${\cal N}=4$ Yang-Mills planar one-loop dilatation operator
\BeisertYB. 
The Bethe ansatz allows for finding (though implicitly) the exact
planar one-loop anomalous dimensions of arbitrary 
local operators in gauge theory. 

On the string theory side of the correspondence, a different integrable
structure has been found in the worldsheet theory \bpr\ and was
related to the gauge theory spin chain description in \dolan.  
Nevertheless, finding the spectrum of string theory in AdS still
remains a challenge. String theory becomes tractable in the
semiclassical regime, when  the quantum numbers of the string states
become large \refs{\gkp}-\refs{\ft}. The energies of certain 
semiclassical soliton solutions with two large angular momenta were
shown to match the anomalous dimensions of operators with large
quantum numbers 
computed to one loop using the XXX spin chain in \BeisertEA\ and to
two loops using
the elliptic spin chain in \SerbanJF. 
In this context, the higher local conserved charges on both sides have
been matched in \ArutyunovRG. 
Moreover, direct agreement has been shown between the continuum limit
of the dilatation operator in the coherent state basis and 
the Hamiltonian of the sigma model \kruc\ without needing the details
of the semiclassical solutions.

Less symmetric spin chains were also analyzed in relation to gauge
theories. The XXZ chain with periodic boundary conditions was related
to pure ${\cal N}=2$ SYM in \XXZ, while the same chain but with
twisted boundary conditions was related in \BECH\ to the $q$-deformation
of the ${\cal N}=4$ SYM. The same gauge theory was shown in \RR\ to be
related to a limit of a 
multi-parametric spin chain which also describes certain
non-supersymmetric theories. 
For related work on integrability in gauge and string theories see
\refs{\othera\otherb\otherc\otherd\othere\otherf\otherg\otherh\otheri
\otherj\otherk\otherl\othero\otherp\otherq\otherr\others
\othert\otheru\otherv\otherw\otherx\othery\otherz\otheraa\otherab\otherac
\otherad\otherae\otheraf\otherag\otherah\otherai\otheraj\otherak
\otheral\otheram\otheran\otherap\otheraq\otherar\otheras-
\otherlast}. 

So far these studies have been limited to finding the eigenvalues 
of the dilatation operator. In this paper we show how to use 
the technology
of integrable spin chains to find three-point functions of
arbitrary scalar operators in ${\cal N}=4$ Yang-Mills and some of its
deformations. \foot{An earlier attempt in this direction \okuyama\
related some of their building blocks to certain open spin
chains. We depart from this standpoint and use only closed spin
chains, which are perhaps more natural from a string theory perspective.}
Even though it is relatively easy to compute three-point functions of
operators at tree level, it becomes increasingly laborious to do so at
higher loops. In fact, even the computation of the tree level 
three-point function coefficient of operators with definite one-loop
scaling dimension is quite hard, because these operators are 
combinations, with occasionally rather complicated coefficients, of
the natural tree level operators.
The spin chain
conveniently identifies a basis of operators with definite scaling
dimension and leads to an algorithm for the computation of the
leading three-point function coefficient. This coefficient turns out
to be related to the expectation value of certain spin chain operators,
a subject which has been extensively studied (see \kbi\ and references
therein). Loop corrections to the three-point function coefficient
may be expressed as the products of expectation values of
operators similar to the spin chain Hamiltonian.

In \S 2 we construct the map between gauge theory and spin chain
correlation functions, emphasizing the spin chain realization of 
arbitrary scalar operators. In \S 3 and \S 4 we review the relevant
details of the algebraic Bethe ansatz and spin chain correlation
functions. We proceed in \S 5 to describe several examples.

At one loop the spin chain description of  various sectors 
of the gauge theory is not restricted to operators
with large quantum numbers. 
The planar perturbative correlation
functions of  these  operators are related to the tree-level
scattering amplitudes of string theory in highly curved AdS.
The limit in which one of the R-charges is large is similar to a
plane wave limit. 
In this regime the corresponding string amplitudes
for massless states have been computed in \jn.
A naive extension of this calculation to massive states
leads to an apparent puzzle: the three-point function involving one
massive string state and two massless states vanishes identically. 
However, one of
the examples we will discuss in \S 5 leads to a nonvanishing result.
Though we will not attempt it here, it would be interesting to clarify
 the relation between the  correlation functions we discuss here
and string theory computations, perhaps along the lines of \YONE. 

\newsec{${\cal N}=4$ Yang-Mills and Spin Chains: the Operator Map}

The computation of planar one-loop anomalous dimensions of 
scalar operators in arbitrary representations of the R-symmetry group 
is conveniently encoded in the diagonalization of the Hamiltonian of
the Heisenberg spin chain with $SO(6)$ symmetry \MIZA,
\eqn\HH{
\Delta={\lambda\over 16\pi^2}
\sum_{i=1}^L K_{i,i+1}+ 2I_{i,i+1} - 2P_{i,i+1} ~~,
}
where $P,~K$ and $I$ are, respectively, the permutation, trace and
identity operators acting on the tensor product of nearest neighbor
spins $i$ and $i+1$ and $L$ is the length of the chain.

The spin chain description also provides a tool for the computation of
various numerical coefficients appearing in correlation functions. 
Such techniques are  particularly efficient  in situations when the
position dependence of the correlation function is known. The 
three-point functions of operators with definite conformal weights
fall in this category.
For operators of  conformal dimension
$\Delta_I, ~\Delta_M$ and $\Delta_K$ the position dependence on the
plane is fixed by (super)conformal invariance to
\eqn\cf{
\langle {\cal O}^I(x) \, {\cal O}^M(y)\,  {\cal O}^K(z)\rangle=
{C^{IMK}(\lambda)\over |x-y|^{\Delta_I+\Delta_M-\Delta_K}
|x-z|^{\Delta_I-\Delta_M+\Delta_K}|y-z|^{-\Delta_I+\Delta_M+\Delta_K}}~~.
}
The OPE coefficients $C^{IMK}(\lambda)$ are in general a series in the
't Hooft parameter
\eqn\eee{
C^{IMK}(\lambda)=c^{IMK}_0(1+\lambda c_1^{IMK}+{\cal O}(\lambda^2))~~.
}
The one-loop correction $c_1^{IMK}$ was partly analyzed in
\okuyama. Before discussing this correction however, it is important
to have an efficient scheme for the computation of the leading
coefficient $c^{IMK}_0$.  

It turns out that the leading OPE coefficients have a simple
description: since they are
given by free field contractions, they can be identified with the 
matrix element of one of the three operators in \cf\ 
with the  in- and out-states corresponding to the other two operators
\eqn\C{
c_0^{IMK}=\langle I| \, {\hat O}^M\,  |J \rangle = 
\langle M| \, {\hat O}^I\,  |J \rangle={\rm etc}~~.
}
Here ${\hat O}^M$ is the representation of ${\cal O}^M$
on the spin chain. There is no canonical way to choose which operators
in \cf\ correspond to the in- and out-states. As we will see later
it is useful to use this freedom to simplify the calculations. For
example, operators with a small number of fields have simple
expressions in terms of the basic spin chain operators and it is
usually convenient to put them in the middle.
We reach the conclusion that to compute $c_0^{IMJ}$ we should
compute the transition induced by some arbitrary operator between two
eigenvectors of scale transformations, in some sector closed under RG
flow.   

An alternative reasoning leads to the conclusion that the leading OPE
coefficients can be expressed in terms of the convolution of three
scalar products of spin chain states. Indeed, given three spin chains
of arbitrary lengths in some real representation it is possible to
define an object similar to a three-string vertex which maps a tensor
product of two states into a third one while taking into account
planar free field contractions. In the case of extremal three-point
functions this construction can be easily extended to spin chains in
complex representations. 

The advantage of this realization of $c_0$ is that it also 
provides a relatively simple way of computing the subleading coefficients
$c_0c_1$ and more generally the coefficients of the logarithmic
corrections to the tree level three-point functions, in terms of the
expectation values of certain local and nonlocal operators. With a
similar goal, the authors of \okuyama\ introduced a map describing
the splitting of a closed spin chain into two open spin chains. The
construction we will  describe at the end of this section bypasses
this step by employing only 
the spin chains describing the  operators whose three-point function we
are interested in computing.

\subsec{The $SU(2)$ Sector}

The holomorphic operators are the easiest ones to analyze in the
context of our discussion. The reason is that the 
dependence of the correlation function on the in- and out-states
becomes easy to track due to R-charge conservation. This becomes
particularly easy for operators belonging to an SU(2) sector of the
theory,
\eqn\operators{
{\cal O}^{i_1\dots i_n}=\Tr[\phi^{i_1}\dots \phi^{i_n}]~~,
}
where $i_k=1,2$ and $\phi^{i}$ are holomorphic fields.

The dilatation operator in this sector is a restriction of \HH\ and
acts on these operators as the negative of the
Hamiltonian of the $SU(2)$ XXX Heisenberg spin
chain, 
\eqn\H{
H=\sum_{i=1}^LH_{i,i+1}
~~~~~~~~
H_{i,i+1}=1-P_{i,i+1}=-(
\sigma^+_i \sigma^-_{i+1}+\sigma^-_i \sigma^+_{i+1}+\sigma^z_i
\sigma^z_{i+1})~~, 
}
with periodic boundary conditions, where
$\sigma^\pm_i={1\over\sqrt{2}}
(\sigma_i^1\pm i\sigma_i^2)$ and $\sigma_i^z$ are
the Pauli matrices acting at the site $i$. 
The eigenvectors of $H$ can be explicitly computed for short
operators, or 
for very long operators, i.e. $L\rightarrow\infty$. In either case
they are linear combinations of \operators\ with coefficients
determined by the Bethe equations, which we will review in \S 3.

Since the operators \operators\ are linearly related to 
the eigenvectors of \H\ which are special configurations of
spins on a chain, it follows that to compute the OPE coefficients we
have to find a representation of some arbitrary operator in terms of
spin chain variables. 
It is not hard to see that there is a qualitative difference between
operators with the same number of holomorphic and antiholomorphic
fields and those in which their number is different. For the first
kind of operators, ordinary
Feynman diagrams imply that the result of the contraction of the 
antiholomorphic fields with fields in the in-state is an operator with
the same number of fields as the original operator and thus it is
described by a state in the original Hilbert space. The operators of
the second kind lead to a change in the number of fields and thus the
resulting state belongs to the space of states of a chain of a
different length. We will treat the two cases separately.

\medskip
\noindent{\it  Operators 
with the same number of $\phi^i$ and ${\bar\phi}_j$}
\smallskip

It is quite easy to find the action of such an operator on 
the eigenvectors of \H\ (or, more generally, on a holomorphic
operator) by analyzing the Feynman rules of the gauge theory. 
Since we
are interested in the planar limit of the gauge theory, let us
consider the operator 
\eqn\middle{
{\cal O}_{j_1\dots j_n}^{i_1\dots i_n}=\Tr[\phi^{i_n}\dots\phi^{i_1}
{\bar\phi}_{j_1}\dots {\bar\phi}_{j_n}]~~,
}
which is schematically represented in figure 1 using the cyclicity of
the trace.

\fig{Graphic representation of the operator \middle.}
{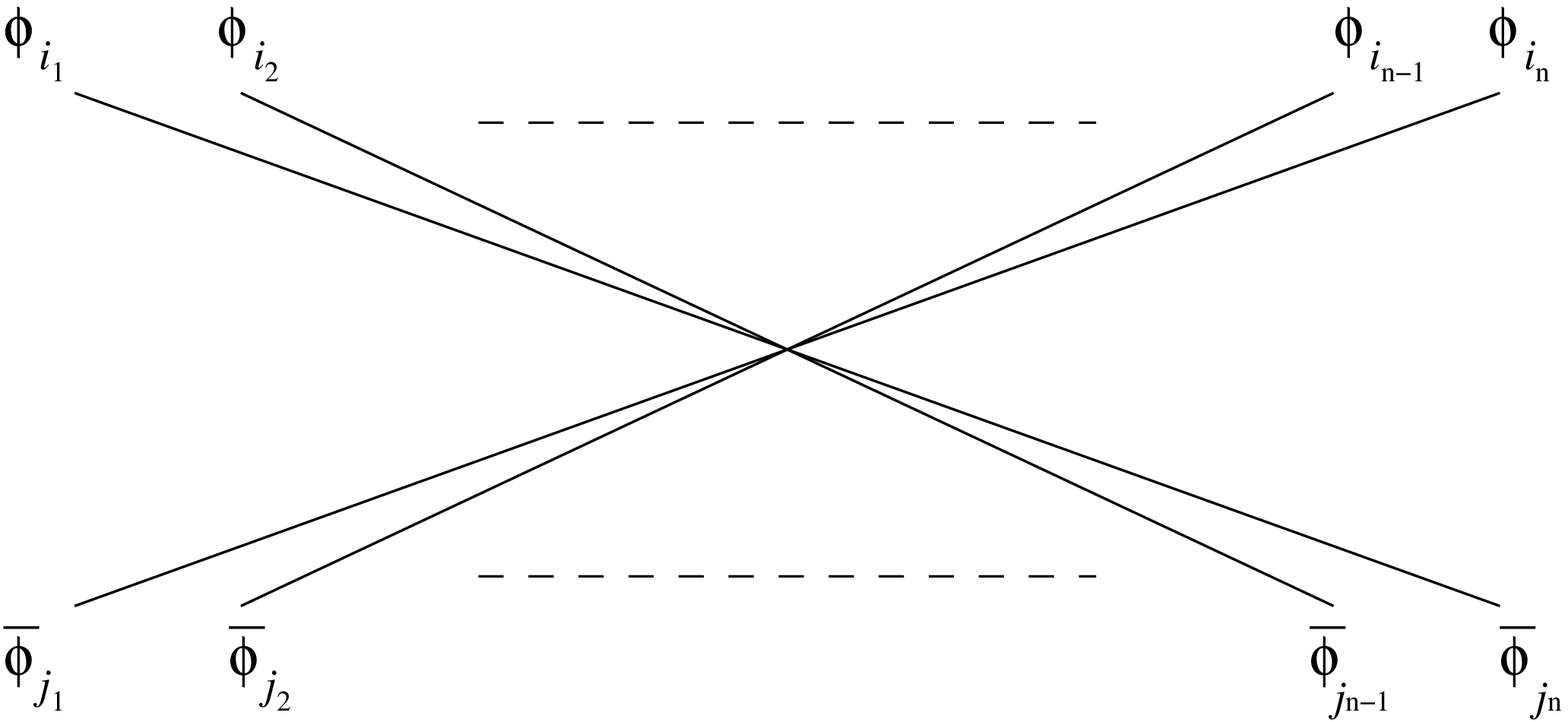}{3.0in}

We see that,
at the planar level, this operator takes $n$ ordered neighboring
fields with $SU(2)$ indices  $(j_1,\dots,j_n)$ and transforms them
into $n$ ordered neighboring fields with indices
$(i_1,\dots,i_n)$. If \middle\ acts on an operator which does not 
contain the sequence of indices $(j_1,\dots,j_n)$ then the
result vanishes identically. On a case by case basis the spin chain
representation of the operator \middle\ can be written quite 
explicitly. 

From figure 1 it is relatively clear how to do this: due to the
cyclicity of the trace we see that the $k$-th field in \middle\ is
transformed into the $2n-k+1$-th field. Depending on which one of 
$\phi^1$, $\phi^2$ or their antiholomorphic versions these fields are,
this transition is realized on the spin chain by one of the $SU(2)$
generators. For example, using conventions compatible with \H\ and 
\S 3, we find that 
\eqn\eee{
\Tr[\phi^1{\bar\phi_1}]~~\longleftrightarrow ~~ \sum_i
{1\over 2}(1+\sigma_i^z)
~~~~~~~~
\Tr[\phi^2{\bar\phi_2}]~~\longleftrightarrow ~~ \sum_i
{1\over 2}(1-\sigma_i^z)~~,
}
which are the number operators for $\phi^1$ and ${\phi}^2$,
respectively. A more involved example is the one containing all
operators with, say, $n$ fields:
\eqn\op{\eqalign{
\Tr\left[\prod_{j=1}^n\left[w_j^1\phi^1+w_{j}^{2}\phi^2\right]
\prod_{i=1}^n\left[v_{n+1-i}^1{\bar\phi}_1+v_{n+1-i}^{2}{\bar\phi}_2
\right]\right]
\longleftrightarrow\cr
\sum_{i=1}^L\prod_{j=i}^{i+n}
\left[u_{i}^{21}\sigma_i^-+u_{i}^{12}\sigma_i^++{u_{i}^{11} \over
2}(1+\sigma_i^z)+{u_{i}^{22} \over 2}(1-\sigma_i^z) \right]~~,
}}
with the coefficients $u$ related to $v$ and $w$ by
\eqn\eee{
u_{i}^{mn}=w_{i}^{m}v_{i}^{n}~~
}
and the site labels defined cyclically $i\simeq i+L$. Computing the
matrix element of this operator between some fixed in- 
and out-states leads to the generating function for the tree
level coefficients
$c_0^{IMJ}$ with fixed $I$ and $J$ (corresponding to the in- and 
out-states) and $M$ of fixed size, but arbitrary $SU(2)$ structure.

\vfil
\eject

\noindent {\it  Operators 
with different number of $\phi^i$ and ${\bar\phi}_j$}
\smallskip

As discussed before, these operators are qualitatively different from
the operators with the same number of holomorphic and antiholomorphic
fields, because their action on the spin chain changes its
length. Indeed, the operator
\eqn\mid{
{\cal O}_{j_1\dots j_n}^{l_1\dots l_m}=\Tr[\phi^{l_m}\dots\phi^{l_1}
{\bar\phi}_{j_1}\dots {\bar\phi}_{j_n}]
}
schematically represented in figure 2, takes $n$ ordered neighboring
fields $(j_1,\dots,j_n)$ and replaces them with $m$ ordered
neighboring fields $(l_1,\dots,l_m)$ changing therefore the number
of fields. Because of this fact, such
operators cannot have an expression similar to \op, but they also must
contain a map between the spaces of states of spin chains of different
lengths.

\fig{Graphic representation of the operator \mid.}
{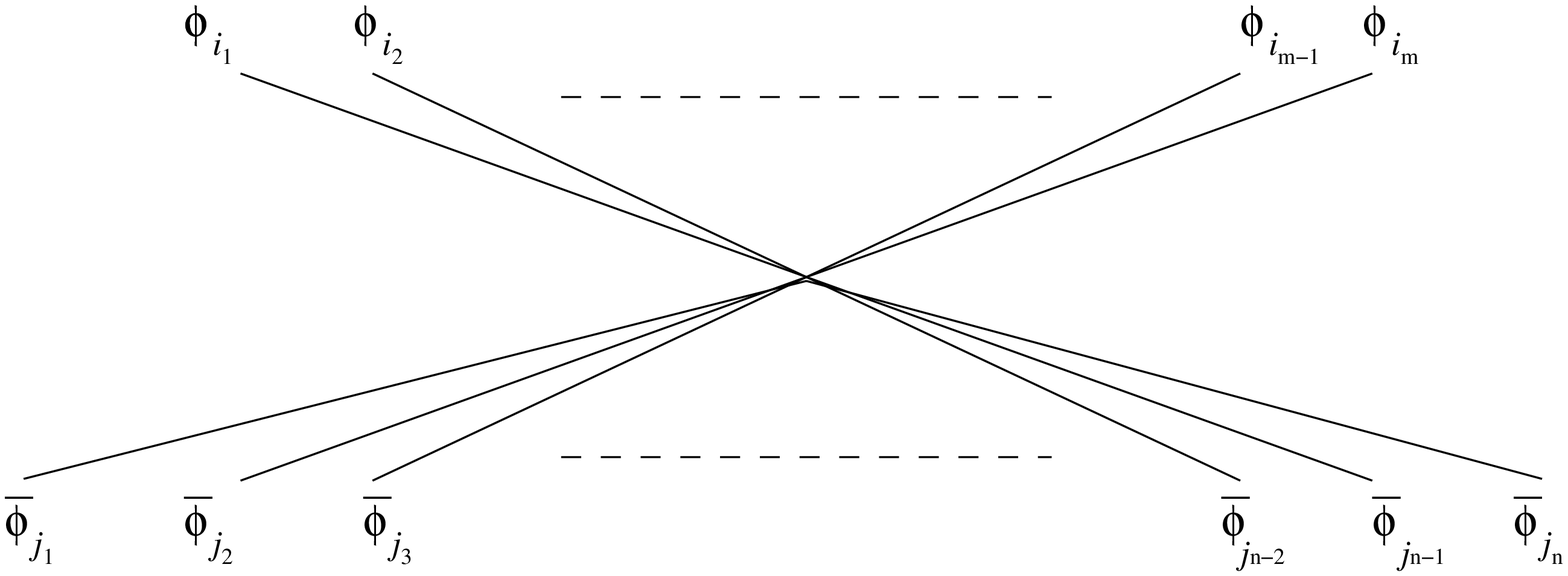}{3.0in}

In the context of our discussion it is relatively easy to construct
this map based on the behavior of the operators \mid. The idea is to
project the in  state on states which contain the sequence
$(j_1,\dots,j_n)$, replace this sequence with $(l_1,\dots,l_m)$ and
sum over all such states. The result is that
we have to compute the expectation value of
\eqn\proj{
{\cal O}_{j_1\dots j_n}^{l_1\dots l_m}
\longleftrightarrow 
\sum_{{k=0 \atop {\rm all}~i_s=1,2}}^{L-1}
|\phi^{i_1}.~.~\phi^{i_{k}}\,
(\phi^{l_1}.~.~\phi^{l_m})\,
\phi^{i_{k+n}}.~.~\phi^{i_{L}}\rangle\langle
{\bar\phi}_{i_1}.~.~{\bar \phi}_{i_{k}}\, 
({\bar\phi}_{j_1}.~.~{\bar\phi}_{j_n})\,
{\bar\phi}_{i_{k+n+1}}.~.~{\bar\phi}_{i_{L}}|
}
where $L$ is the length of the chain corresponding to the in-state,
the sum is over all $i_1=1,2;~\dots i_{k-1}=1,2;~i_{k+n+1}=1,2;\dots
i_N=1,2;$ and
for $k\ge L-n$ the states are defined using the cyclicity of the trace.
Each of the two states appearing in every term in the sum above  can be
obtained by acting with $SU(2)$ generators on the out and in 
ground states of the spin chains of length $L$ and $L+m-n$,
respectively, and must be unit normalized. 
Here $|\phi^{i_1} \dots  \phi^{i_n} \rangle= |\phi^{i_1}\rangle
\otimes \dots \otimes |\phi^{i_n} 
\rangle$ where  $|\phi^1\rangle=|0\rangle$  and $|\phi^2
\rangle=\sigma^- |0\rangle.$ 

It is worth mentioning that the representation \proj\ extends to
operators 
with the same number of holomorphic and antiholomorphic
fields \middle\ by just picking $m=n$. Depending on the context, it 
may be that one of the two representations \op\ or \proj\ is more
convenient. Clearly, \proj\ leads to OPE coefficients involving
only scalar products which may be better-suited for a general analysis.

An additional useful point is that the equation \proj\ can be
conveniently written in terms of the shift operator along the
chain, $\tau(0)$,
\eqn\eee{
\tau(0)|\phi^{i_1}\phi^{i_2}\dots\phi^{i_{L}}\rangle=
|\phi^{i_L}\phi^{i_1}\dots\phi^{i_{L-1}}\rangle~~.
}
For the purpose of computing correlation functions of gauge invariant
operators (represented by eigenvectors of the spin chain transfer
matrix) this is particularly useful because these operators are
invariant under such shifts. Thus, using
\eqn\projshift{
{\cal O}_{j_1\dots j_n}^{l_1\dots l_m}~~
\longleftrightarrow ~~
\sum_{{k=0 \atop {\rm all}~i_s=1,2}}^{L-1}
\tau(0)^k|
(\phi^{l_1}.~.~\phi^{l_m})\,
\phi^{i_{n+1}}.~.~\phi^{i_{L}}\rangle\langle
({\bar\phi}_{j_1}.~.~{\bar\phi}_{j_n})\,
{\bar\phi}_{i_{n+1}}.~.~{\bar\phi}_{i_{L}}|\tau(0)^k
}
the summation over $k$ leads only to multiplication by the length of
the chain. 

\subsec{The $SO(6)$ Sector}

The discussion in the previous subsection extends, with some
modification, to operators in the $SO(6)$ sector, i.e. all scalar
operators in the theory. The main difference is that, unlike in the
$SU(2)$ case, the fields transform in a real representation and
therefore all fields in the operator whose matrix element we want to
compute can have a nontrivial contraction with the fields in the in-
and out-states. The only constraint comes from its trace structure.

As before, given an in- and out-state with $k_I$ and $k_J$ fields
respectively, an operator with $k_M$ fields will have a nontrivial
matrix element between the given in- and out-states if
\eqn\nonzero{
k_I=m+n~~~~~~~~k_J=m+p~~~~~~~~k_M=n+p~~.
}
Similarly to the $SU(2)$ case, it is useful to make a distinction
between the cases $n=p$ (or $k_I=k_J$) and $n\ne p$ (or $k_I\ne k_J$).
In the first case we will have again
two apparently different expressions ${\hat O}^M$ which may be
useful in different contexts.

If $n=p$ the operator ${\hat O}^M$ can be schematically represented as
in figure 1, except that now we have to sum over all possible ways of
choosing $n$ neighboring fields to be contracted with the in-state.
The operator
\eqn\eee{
{\cal O}^M=\Tr[\phi^{i_1}\dots \phi^{i_{2n}}]
}
is represented on a spin chain by
\eqn\eee{
\sum_{j=1}^L \sum_{k=1}^{2n}(E_j)^{i_{2n+1-k}}_{i_k}\dots 
(E_{j+n-1})^{i_{k+n}}_{i_{k+n-1}}=\sum_{i=1}^L \tau(0)^i
\sum_{k=1}^{2n}(E_1)^{i_{2n+1-k}}_{i_k}\dots 
(E_{n})^{i_{k+n}}_{i_{k+n-1}}\tau(0)^{-i},
}
where $(E_n)_i^j$ are the
generators of the general linear group 
\eqn\eee{
((E_n)^i_j)_k^l = \delta^i_k\delta^l_j
}
with the index $n$ labeling the site and all indices are defined
cyclically 
\eqn\eee{
i+L\simeq i~~~~~~~~i_{k+2n}\simeq i_k~~.
}

If $n\ne p$ the operator ${\hat O}^M$ can be schematically represented
as in figure 2, with the same provision that we have to sum over all
possibilities of choosing $n$ neighboring fields to be contracted
with the fields of the in-state. The analog of equation \projshift\
is
\eqn\projsix{
{\cal O}^{n+p}
\longleftrightarrow \!\!\!\!\!\!\!\
\sum_{{k=0\atop {\rm all}~i_s=1\dots 6}}^{L-1}
\!\!\!\!\!\!
\tau(0)^k|
(\phi^{l_{s+1}}.~.~\phi^{l_{s+p}})\,
\phi^{i_{k+n+1}}.~.~\phi^{i_{L}}\rangle\langle
({\phi}^{l_{s+p+1}}.~.~{\phi}^{l_{s+n+p}})\,
{\phi}^{i_{k+n+1}}.~.~{\phi}^{i_{L}}|\tau(0)^k
}
where, as before, all indices are cyclically defined
\eqn\eee{
i+L\simeq i~~~~~~~~l_{s+n+p}\simeq l_s~~.
}

It is certainly possible to extend this construction to operators
containing fermions. Perhaps the simplest such sector is the one
containing two scalars and one fermion (the gaugino in an ${\cal N}=1$
language), which has a manifest $SU(1|2)$ symmetry. The
extension to the full theory (the full $SU(4|4)$ spin chain) is
complicated by the fact that the Feynman diagram interpretation of the
dilatation operator is not immediately obvious. We
will not attempt to discuss this here and leave it for future work.

An important observation is that the technique used to
construct the matrix elements of operators with $n\ne p$ and different
number of holomorphic and antiholomorphic fields can be also used to
describe the mixing of single- and double-trace operators. Indeed,
equations analogous to \proj\ and \projsix\ represent maps between the
space of states of a single chain and the tensor product of the spaces
of states of two different chains of different lengths. 
Following the logic outlined before, the leading OPE coefficient 
describing the transition between a single-trace and a double-trace operator
is given by the matrix element of 
\eqn\tree{
\sum_{{s=0\atop {\rm all}~i_n=1\dots 6}}^L
|\phi^{i_{s+L_1+1}}\dots\phi^{i_{s+L_2}}\rangle_{{}_{{\rm mod}\,L_2}}
\otimes
|\phi^{i_{s+1}}\dots\phi^{i_{s+L_1}}\rangle_{{}_{{\rm mod}\,L_1}}\langle
{\bar\phi}_{i_1}\dots
({\bar\phi}_{i_{s+1}}\dots
{\bar\phi}_{i_{s+L_1}})
\dots{\bar\phi}_{i_{L}}|
}
between an eigenstate of a chain of length $L$ and the tensor product
of two states belonging to chains of lengths $L_1$ and $L_2=L-L_1$,
respectively. In the same spirit, the leading OPE coefficient for three
operators described by spin chains in a real representation (such as
the vector representation of $SO(6)$) is given by the expectation
value of 
\eqn\treesix{
\eqalign{
&\sum_{s_1=0}^{L_1-1}\sum_{s_2=0}^{L_2-1}\sum_{s_3=0}^{L_3-1}
\sum_{{\rm all}~i,j}
\tau(0)_{L_1}^{s_1}|(\phi^{i_{1}}\dots\phi^{i_{m}})(\phi^{i_{m+1}}\dots
\phi^{i_{L_1}})\rangle \otimes 
\cr
&\tau(0)_{L_2}^{s_2}|(\phi^{i_{1}}\dots\phi^{i_{m}})(\phi^{j_{m+1}}\dots
\phi^{j_{L_2}}) \rangle 
\langle
({\phi}_{i_{m+1}}\dots {\phi}_{i_{L_1}})
({\phi}_{j_{m+1}}\dots{\phi}_{j_{L_2}})|\tau(0)^{s_3}_{L_3}
}
}
between an eigenstate of a chain of length $L_1$ and the tensor product
of two states belonging to chains of lengths $L_2$ and $L_3$ obeying
a relation similar to \nonzero.

\subsec{The One-Loop Correction}

The form given by \treesix\ from the tree level OPE coefficient lends
itself to higher order calculations as well. For the sake of simplicity
we will exemplify this at the one-loop level, but the extension to
higher loops is conceptually relatively clear.

There are two different types of Feynman diagrams contributing to a
one-loop three-point function: 1) the loop involves fields belonging
to only two of the three operators and  2) the four fields belong to
three operators. Since only F-terms contribute nontrivially to 
three-point functions \DHoker, in both cases the R-symmetry index
structure is given by the  one-loop Hamiltonian \HH, except that the
sites it acts on may belong to different chains
\eqn\eee{
{\tilde H}:V^{(a)}_i\otimes V^{(b)}_{j}\rightarrow V^{(c)}_k\otimes 
V^{(d)}_{l}
~~~~~~~~~~~~
{\tilde H}=H~~,
}
where $a,\,b,\,c,\,d$ label the chain while $i,\,j,\,k,\,l$ label the
sites. The different types of diagrams have a different position
dependence. The first 
type leads to the same position dependence as that of the one-loop
corrections to two-point functions, $\ln |x_i-x_j|^2$,  
while the second type leads to the position dependence
typical for one-loop corrections to three-point functions, $\ln
|x_i-x_j|^2|x_i-x_k|^2/|x_j-x_k|^2$. Identifying the labels of 
the positions $x_i$ and the length of the operators $L_i$, 
the coefficient of $\ln |x_1-x_2|^2$ in a generic three-point function
is given by the expectation value of
\eqn\twosix{
\eqalign{
&\sum_{s_1=0}^{L_1-1}\sum_{s_2=0}^{L_2-1}\sum_{s_3=0}^{L_3-1}
\sum_{{\rm all}~i,j}
\tau(0)_{L_1}^{s_1}\otimes\tau(0)_{L_2}^{s_2}
H_{i_ki_{k+1}}^{j_kj_{k+1}}
|(\phi^{i_{1}}
\dots\phi^{i_{k}}\phi^{i_{k+1}}\dots\phi^{i_{m}})(\phi^{i_{m+1}}\dots
\phi^{i_{L_1}})\rangle 
\cr
&\otimes
|(\phi^{i_{1}}\dots\phi^{j_{k}}\phi^{j_{k+1}}
\dots\phi^{i_{m}})(\phi^{j_{m+1}}\dots
\phi^{j_{L_2}}) \rangle
\langle
({\phi}_{i_{m+1}}\dots {\phi}_{i_{L_1}})
({\phi}_{j_{m+1}}\dots{\phi}_{j_{L_2}})|~\tau(0)^{s_3}_{L_3}~~.
}
}
Similarly, the coefficient of $\ln
|x_3-x_1|^2|x_3-x_2|^2/|x_1-x_2|^2$ is given by the expectation value
of 
\eqn\threesix{
\eqalign{
&\sum_{s_1=0}^{L_1-1}\sum_{s_2=0}^{L_2-1}\sum_{s_3=0}^{L_3-1}
\sum_{{\rm all}~i,j,i',j'}
\tau(0)_{L_1}^{s_1}
\otimes\tau(0)_{L_2}^{s_2}~\left[
H_{i_{L_1}j_{m+1}}^{i'_{L_1}j'_{m+1}}I_{i_{m+1}j_{L_2}}^{i'_{m+1}
j'_{L_2}}
\right.
+\left.
I_{i_{L_1}j_{m+1}}^{i'_{L_1}j'_{m+1}}H_{i_{m+1}j_{L_2}}^{i'_{m+1}j'_{L_2}}\right]
\cr
&|(\phi^{i_{1}}
\dots\phi^{i_{m}})(\phi^{i_{m+1}}\dots
\phi^{i_{L_1}})\rangle 
\otimes
|(\phi^{i_{1}}
\dots\phi^{i_{m}})(\phi^{j_{m+1}}\dots
\phi^{j_{L_2}}) \rangle
\cr
&\langle
({\phi}_{i'_{m+1}}{\phi}_{i_{m+2}}\dots {\phi}_{i_{L_1-1}}
{\phi}_{i'_{L_1}})
({\phi}_{j'_{m+1}}{\phi}_{j_{m+2}}\dots{\phi}_{j_{L_2-1}}
{\phi}_{j'_{L_2}})|~\tau(0)^{s_3}_{L_3}~~.
}
}
The generalization to the other terms appearing in the one-loop
corrections to three-point functions is now obvious. For the case of
extremal three-point functions\foot{It would be interesting to extend
this analysis to higher point functions,
where extremality is likely to simplify the calculation.
In the context of AdS/CFT extremal and near-extremal 
$n$-point functions of chiral operators have been discussed in \DFMMR\
and \DEFP\ respectively.} there are yet two more 
contributions due to the fact that the possible Feynman diagrams can
have more general topologies (see e.g. \GMRII). 
These new contributions are similar to
\twosix\ and \threesix\ and can be easily inferred from Feynman diagrams.

We will now proceed to describe the calculation of matrix elements
of some of the spin chain operators constructed in this section. For
the sake of simplicity we concentrate on the $SU(2)$ sector, but
some of the formulae extend to larger collections of operators. Though
we will mostly be concerned with the representation \op\ of gauge theory
operators on the spin chain, the general expressions 
which we will review in \S 4 are all the
necessary ingredients in the calculation of the expectation values of
\projshift, \projsix, \tree, \treesix, \twosix\ and \threesix\ as well.

\newsec{Algebraic Bethe Ansatz: a Quick Review}

The $SU(2)$ sectors of ${\cal N}=4$ SYM and some of its deformations
are described by the spin 1/2 XXX and XXZ spin chains. 
The Hamiltonian of the XXZ chain is
\eqn\aaa{
H=
\sum_{i=1}^L  (\sigma^x_i \sigma^x_{i+1}+
\sigma^y_i \sigma^y_{i+1}+ \Delta(\sigma^z_i \sigma^z_{i+1}-1))~~,
}
where $\sigma^a_i, ~a=x,y,z$ are the Pauli matrices acting at the site
$i$ and for $\Delta=1$ we recover the XXX chain. This model can be
solved by the Bethe Ansatz which we will now briefly review following
\kbi,\faddeev. 

The starting point is the $R$-matrix
which satisfies the Yang-Baxter equations. In our cases, the relevant
solutions are
\eqn\aaa{
R(\lambda)=-i
\pmatrix{\alpha(\lambda)&0&0&0\cr
0&\beta(\lambda)&\gamma(\lambda)&0\cr
0&\gamma(\lambda)&\beta(\lambda)&0\cr
0&0&0&\alpha(\lambda)
}~~,
}
where $\lambda$ is the spectral parameter or rapidity, while
\eqn\aaa{
\alpha(\lambda)=i{\sinh(\lambda+2 i \eta)\over\sinh 2 i \eta}~~,~~~~~
\beta(\lambda)=i{\sinh\, \lambda\over\sinh 2 i \eta}~~,~~~~~
\gamma(\lambda)=i
}
for the XXZ chain and
\eqn\aaa{
\alpha(\lambda)=\lambda+{i}~~,~~~~~
\beta(\lambda)={\lambda}~~,~~~~~
\gamma(\lambda)=i
}
for the XXX chain. The $R$-matrix is normalized such
that $R(0)$ is the permutation operator. 

The monodromy matrix is
constructed out of the $R$ matrix as
\eqn\monod{
T^{a_L;k_L\dots k_1}_{\; b_1;n_L\dots n_1}(\lambda)=
R^{a_Lk_L}_{b_Ln_L}
R^{b_Lk_{L-1}}_{b_{L-1}n_{L-1}}
\dots R^{b_{2}k_1}_{b_1n_1}(\lambda)
\equiv
\pmatrix{ A^{k_L\dots k_1}_{n_L\dots n_1}(\lambda)& B^{k_L\dots
k_1}_{n_L\dots n_1}(\lambda) \cr 
C^{k_L\dots k_1}_{n_L\dots n_1}(\lambda) & D^{k_L\dots k_1}_{n_L\dots
n_1}(\lambda) },
}
where the indices $(k_L,\dots,k_1)$ and $(n_L,\dots,n_1)$ correspond
to the sites of the chain while $a_L$ and $b_1$ correspond to an
auxiliary space of the same dimension as the dimension of the vector
space on the sites of the chain. As a consequence of the Yang-Baxter
equations, the monodromy matrix satisfies
\eqn\RTT{
R(\lambda-\mu) (T(\lambda)  \otimes T(\mu))=
(T(\mu)  \otimes T(\lambda))  R(\lambda-\mu)~~,
}
where $R$ and $\otimes$ act in the auxiliary space. Examining the
various entries 
of this matrix equation it is relatively easy to find that the 
off-diagonal entries of $T$ act almost as creation and annihilation
operators with respect to the diagonal entries. The relevant 
matrix elements of \RTT\ can be written as
\eqn\algebra{\eqalign{
[B(\lambda),B(\mu)]=[C(\lambda),C(\mu)]=0~~, \cr
[B(\lambda),C(\mu)]=g(\lambda,\mu) (D(\lambda) A(\mu)-D(\mu)
A(\lambda))~~, \cr 
D(\mu) B(\lambda)=f(\lambda,\mu) B(\lambda) D(\mu)+g(\mu,\lambda)
B(\mu) D(\lambda)~~, \cr 
A(\mu) B(\lambda)=f(\mu,\lambda) B(\lambda) A(\mu)+g(\lambda,\mu)
B(\mu) A(\lambda)~~, 
}}
where 
\eqn\fgdef{
f(\lambda,\mu)\equiv f(\lambda-\mu)={\alpha(\mu-\lambda)\over
\beta(\mu-\lambda)}~~,~~~~~ 
g(\lambda,\mu)\equiv g(\lambda-\mu)=-{\gamma(\lambda-\mu) \over
\beta(\lambda-\mu)}~~. 
}

A consequence of the Yang-Baxter  equations is that the transfer
matrix, which is the trace of the monodromy matrix in the
auxiliary space 
\eqn\tr{
\tau(\lambda)=\Tr_{aux}\, T(\lambda)=A(\lambda)+D(\lambda)~~,
}
commutes with itself at arbitrary values of spectral parameters.
Thus, if the Hamiltonian is among the Taylor coefficients of
$\tau(\lambda)$,  the transfer matrix is the generating functional of
an infinite number of mutually commuting conserved charges. 
The Hamiltonian is constructed out of the transfer matrix as
\eqn\aaa{
 H^{XXZ}= i \sin 2 \eta \,{d \over d \lambda}
\ln\tau_{XXZ}(\lambda)\Big|_{\lambda=i \eta}-  L \cos 2 \eta~~~~~~
H^{XXX}= {d \over d \lambda}
\ln\tau_{XXX}(\lambda)\Big|_{\lambda=0}-  L 
}
and $\Delta=\cos\eta$. 

To construct the  eigenstates of the transfer matrix and the Hamiltonian
one starts with the (pseudo)vacuum $|0 \rangle$ which satisfies
\eqn\vac{
A(\lambda)|0\rangle=\tilde{a}(\lambda)|0\rangle,~~~
D(\lambda)|0\rangle=\tilde{d}(\lambda)|0\rangle,~~~
C(\lambda)|0\rangle=0,
}
where 
\eqn\zen{
{\tilde a}(\lambda)=(-i\alpha(\lambda))^L~~,~~~
{\tilde d}(\lambda)=(-i\beta(\lambda))^L
}
for both the XXX and the XXZ chain.
With this starting point the natural ansatz for the eigenvectors 
of the transfer matrix is
\eqn\eve{
|\Psi_N(\{ \lambda\})\rangle=
\prod_{i=1}^N B(\lambda_i) |0\rangle,
~~~~N=0,1, \dots,L
}
where the arguments $\lambda_i$ are determined by the requirement that
\eve\ be indeed an eivenvector of \tr. Using the fact that \eve\ is
symmetric in $\lambda_i$ it is possible to find the action of the
operators $A(\mu),~D(\mu)$ and $C(\mu)$ 
on $|\Psi_N(\{ \lambda\})\rangle$
\eqn\adc{\eqalign{
A(\mu) \prod_{j=1}^N B(\lambda_j) |0 \rangle=
\Lambda(\mu) \prod_{j=1}^N B(\lambda_j) |0\rangle+
\sum_{n=1}^N \Lambda_n B(\mu) \prod_{j=1,j\neq n}^N
B(\lambda_j)|0\rangle~~, 
\cr
D(\mu) \prod_{j=1}^N B(\lambda_j) |0 \rangle=
\tilde{\Lambda}(\mu) \prod_{j=1}^N B(\lambda_j) |0\rangle+
\sum_{n=1}^N \tilde{\Lambda}_n B(\mu) \prod_{j=1,j\neq n}^N
B(\lambda_j)|0\rangle~~, 
\cr
C(\mu) \prod_{j=1}^N B(\lambda_j) |0 \rangle=
\sum_{n=1}^N M_n \prod_{j=1 j \neq n}^N B(\lambda_j)|0 \rangle+
\sum_{k>n} M_{kn} B(\mu) \prod_{j=1,j \neq k, n} B(\lambda_j)|0\rangle
}}
where
\eqn\aaa{\eqalign{
&\Lambda(\mu)=a(\mu) \prod_{j=1}^N f(\mu,\lambda_j), ~~~
\Lambda_n=a(\lambda_n) g(\lambda_n,\mu)\prod_{j=1,j\neq n}
f(\lambda_n,\lambda_j)~~, 
\cr
&\tilde{\Lambda}(\mu)=d(\mu) \prod_{j=1}^N f(\lambda_j,\mu),~~~
\tilde{\Lambda}_n=d(\lambda_n) g(\mu,\lambda_n)\prod_{j=1,j\neq n}
f(\lambda_j,\lambda_n)~~, 
\cr
&M_n=g(\mu,\lambda_n) a(\mu) d(\lambda_n) \prod_{j \neq n}
f(\lambda_j,\lambda_n) 
f(\mu,\lambda_j)+g(\lambda_n,\mu) a(\lambda_n) d(\mu) \prod_{j\neq n}
f(\lambda_j,\mu) 
f(\lambda_n,\lambda_j)~~,
\cr
&M_{kn}=d(\lambda_k) a(\lambda_n) g(\mu,\lambda_k) g(\lambda_n,\mu)
f(\lambda_n,\lambda_k) 
\prod_{j\neq k,n}f(\lambda_j,\lambda_k) f(\lambda_n,\lambda_j)
\cr
&~~~~~~~~~~~~~~~~~~~~
+d(\lambda_n) a(\lambda_k) g(\mu,\lambda_n) g(\lambda_k,\mu)
f(\lambda_k,\lambda_n) 
\prod_{j\neq k,n}f(\lambda_j,\lambda_n) f(\lambda_k,\lambda_j)~~.
}}
and
\eqn\shifted{
a(\lambda)={\tilde a}(\lambda-k)
~~~~~~~~
d(\lambda)={\tilde d}(\lambda-k)
}
with an arbitrary constant $k$. This shift does not affect
any of the functions $f$ and $g$ above because they depend only on
differences of rapidities. The standard choices, which make the
equations most symmetric, are $k=i/2$ and $k=i\eta$ for the XXX and XXZ
chains respectively.

It thus follows that $|\Psi_N(\{ \lambda\})\rangle$ is an eigenvector
of \tr\ if $\lambda_i$ satisfy the Bethe equations
\eqn\beq{
\prod_{k=1, k\neq j}^N {f(\lambda_j,\lambda_k) \over
f(\lambda_k,\lambda_j)}=  {d(\lambda_j) 
\over a(\lambda_j)}~~~~~~~~~(\forall)~~j=1,\dots,N~~,
}
and the corresponding eigenvalues of the transfer matrix are 
\eqn\aaa{
\tau(\mu) |\Psi_N(\{ \lambda\})\rangle=(\Lambda(\mu)+{\tilde\Lambda}(\mu))
|\Psi_N(\{ \lambda\})\rangle~~.
}

An additional important constraint comes from the requirement that the
eigenvectors correspond to gauge invariant operators. This translates
into the requirement that they be invariant under shifts along the
chain. Using the fact that $R(0)$ is the permutation operator, it
follows that $\tau(0)$ generates such shifts. Thus, the last
constraint on the eigenvectors is
\eqn\aaa{
\Lambda(\textstyle{i\over 2})+{\tilde\Lambda}(\textstyle{i\over 2})=1~~.
}

\newsec{Computation of Correlation Functions}

To compute the various expectation values and scalar products derived
in \S 2 it is necessary to find a realization of the spin operators in
terms of the entries of the monodromy matrix. 
Such a representation
makes it possible to use the relations \algebra\ and \adc\ to perform
computations.

\subsec{Inverse Scattering Method}

The inverse scattering method expresses the local spin operators in
terms of the matrix elements of the transfer matrix. The idea of this
method is relatively easy to understand. It is based on the fact
that there exists a value for the rapidity $\lambda$ for which the
$R$-matrix becomes the permutation operator. Using its properties we
can see that
\eqn\aaa{
T(0)=P_{0L}P_{0L-1}\dots P_{01}=P_{01}P_{1L}P_{1L-1}\dots P_{12}=
P_{01}P_{12}P_{23}\dots P_{L-1L}=P_{01}\,\tau(0)~~.
}
Since $P_{01}=\sum_{a=1}^3\sigma_0^a\otimes\sigma_1^a$,
we find \GohmannAV\ that
\eqn\spin{
\eqalign{
\sigma_1^+=C(0)\tau(0)^{L-1}~~~~~~~&~~~~~~~\sigma_1^-=B(0)\tau(0)^{L-1}
\cr
{1\over 2}(1+\sigma^z_1)=A(0)\tau(0)^{L-1}
~~~~~~~&~~~~~~~
{1\over 2}(1-\sigma^z_1)=D(0)\tau(0)^{L-1}~~,
}
} 
where we used the fact that translation by the length of the chain
acts as the identity operator\foot{ Note that this expression holds
for all models based on an $R$-matrix satisfying $R(0)=P$ and
monodromy matrix constructed as in \monod. Thus, the
invariant meaning of the arguments of $A,~B,~C,~D$ and $\tau$ in
equation \spin\ is the value of the spectral parameter for which the
$R$-matrix becomes the permutation operator. 
It will occasionally be convenient to shift
the rapidity, as in equation \shifted.}.   
From here it is easy to find the spin operators at the site $i$, using
the shift operator $\tau(0)$: 
\eqn\shi{
O_i=\tau(0)^{i}O_1\tau(0)^{-i}~~.
}

It is however not hard to see that, using these expressions, the
equations \adc\ lead to apparently singular results if we act on some
state with two or more operators of the same rapidity. Indeed, both
$f(0)$ and $g(0)$ are singular, and going back to equation \RTT\
does not yield the necessary commutation relations. Using
the explicit form of the spin operators one can easily
convince oneself that the scalar products are completely finite. Thus,
for computational purposes, it is necessary to regularize \spin. 
A convenient regularization is provided by the anisotropic chains,
which are based on the same $R$-matrix as the isotropic one, except
that the $R$-matrices building the monodromy matrix are evaluated at 
site-dependent shifted rapidities
\eqn\anis{
T(\lambda)=
R_{0L}
(\lambda-\xi_L) R_{0L-1}
(\lambda-\xi_{L-1}) 
\dots R_{01}
(\lambda-\xi_1)~~,
}
and the vacuum state \vac\ and the vacuum energies \zen\ are
modified appropriately.

Though the arguments are more involved \kitaninedet , 
the expressions for the spin
operators in terms of the entries of the monodromy matrix are quite
similar to \spin\ and \shi:
\eqn\ism{\eqalign{
\sigma_i^{-}&=\{ \prod_{a=1}^{i-1} (A+D)(\xi_a)\}  B(\xi_i) \{
\prod_{a=i+1}^{L} (A+D)(\xi_a)\}~~, 
\cr
\sigma_i^{+}&=\{ \prod_{a=1}^{i-1} (A+D)(\xi_a)\}  C(\xi_i) \{
\prod_{a=i+1}^{L} (A+D)(\xi_a)\}~~, 
\cr
{1\over 2}(1+\sigma^{z}_i)&=\{ \prod_{a=1}^{i-1} (A+D)(\xi_a)\}  
 A(\xi_i) \{ \prod_{a=i+1}^{L} (A+D)(\xi_a)\}~~,
\cr
{1\over 2}(1-\sigma^{z}_i)&=\{ \prod_{a=1}^{i-1} (A+D)(\xi_a)\}  
 D(\xi_i) \{ \prod_{a=i+1}^{L} (A+D)(\xi_a)\}~~.
}}
According to \GohmannAV\ similar formulae hold for superalgebras,
which generalizes our discussion to some sectors of gauge theory
containing fermionic operators.

\subsec{Scalar Product}

As we have seen in \S 2, the calculation of correlation functions of
gauge 
invariant operators reduces at the level of the spin chain to
the calculation of the scalar products of states constructed out of 
$B$ and $C$ operators
\eqn\scalar{
S_N=
\langle0| \prod_{j=1}^N C(\lambda_j^C)\prod_{k=1}^N 
B(\lambda_k^B)|0\rangle~~,
}
where $\lambda_j$ are kept arbitrary (i.e. not necessarily satisfying the
Bethe equations).  There is a large amount of literature devoted
to the computation of such inner products, part of it described in
detail in \kbi . One of the main results is that \scalar\ is given by
\eqn\aaa{\eqalign{
S_N=&\prod_{j<k} g(\lambda_j^C,\lambda_k^C) g(\lambda_k^B,\lambda_j^B)
\sum {\rm sign}(P_C) {\rm sign}(P_B) \prod_{j, k}
h(\lambda_j^{AB},\lambda_k^{DC}) 
\prod_{l,m} h(\lambda_l^{AC},\lambda_m^{DB})
\cr
&\times \prod_{l,k} h(\lambda_l^{AC},\lambda_k^{DC})
\prod_{j,m} h(\lambda_j^{AB},\lambda_m^{DB})
\det(M_{DC}^{AB}) \det(M_{DB}^{AC})~~,
}}
where $P_C$ and $P_B$ are the permutations 
$\{\lambda_1^{AC}, \dots,\lambda_n^{AC},\lambda_1^{DC},\dots,\lambda_{N-n}^{DC} \}$ of
$\{\lambda_1^C,\dots,\lambda_N^C\}$
and 
$\{\lambda_1^{DB}, \dots,\lambda_n^{DB},\lambda_1^{AB},\dots,\lambda_{N-n}^{AB}\}$
of  $\{\lambda_1^B,\dots,\lambda_N^B\}$
respectively, and
\eqn\aaa{
(M_{DC}^{AB})_{jk}=t(\lambda_j^{AB},\lambda_k^{DC}) d(\lambda_k^{DC})
a(\lambda_j^{AB}), ~~~~t(\lambda,\mu)={g(\lambda,\mu)^2 \over
f(\lambda,\mu)}, 
~~~~h(\mu,\lambda)={f(\mu,\lambda) \over g(\mu,\lambda)}.
} 
This result is derived using the commutation relations \algebra\ as
well as various properties of rational functions. When the in-state 
is an energy eigenstate (i.e. $\{\lambda^B\}$ is a solution of the
Bethe equations \beq) \scalar\ simplifies considerably and takes the
following form \kitaninedet\
\eqn\sp{
S_N(\{\lambda^C\},\{\lambda^B\})={
\prod_{i} d(\lambda^C_i)\prod_{a} d(\lambda^B_a)
\det H(\{\lambda^C_i\},\{\lambda^B_k\})\over \prod_{j>k} 
\varphi(\lambda^C_k-\lambda^C_j)
\prod_{l<m} \varphi(\lambda^B_m-\lambda^B_l)}
}
where
\eqn\hmat{
H_{ab}={\varphi(\eta)\over \varphi(\lambda^B_a-\lambda^C_b)}\left[
{a(\lambda^C_b)\over d(\lambda^C_b)}\prod_{m\ne
a}\varphi(\lambda^B_m-\lambda^C_b + \eta_0) -  
\prod_{m\ne a}\varphi(\lambda^B_m-\lambda^C_b - \eta_0)
\right]~~.
}
The function $\varphi(\lambda)$ equals $\lambda$ and  
$\sinh\lambda$ for the XXX and XXZ chains, respectively, while $\eta_0$
equals $i$ and $2i\eta$ for the XXX and XXZ chains, respectively.

The entries of the matrix $H$ have potential singularities if some
rapidity $\lambda_k^B$ of the out-state approaches one of the rapidities
$\lambda_m^C$ of the in-state. This apparent problem is however cured by
the fact that, in this limit, the square bracket in \hmat\ turns out
to vanish due to the Bethe equations \beq. This type of scalar
product is sufficient for the calculation of three-point function
coefficients and will be useful in the examples we will discuss in \S 5.
In the limit in which the two states become identical, \sp\ becomes
the square of the norm of the Bethe eigenstates given by
the Gaudin formula proven in \KORnorm 
\eqn\norm{
\langle0| \prod_{j=1}^N C(\lambda_j)\prod_{k=1}^N 
B(\lambda_k)|0\rangle=\prod_{i} d(\lambda_i) d(\lambda_i)
\prod_{a\neq b} f(\lambda_a,\lambda_b)
\det \Phi'(\{\lambda\})~~,
}
where $\Phi'$ is an $N \times N$ matrix with the elements
\eqn\phider{
\Phi'_{ab}=-{\partial \over \partial \lambda_b}
\log \left({a(\lambda_a) \over d(\lambda_a)} 
\prod_{k=1, k\neq a}^N {f(\lambda_a,\lambda_k) \over 
f(\lambda_k,\lambda_a)}
\right)~~.
}
The explicit form of these matrix elements is not too complicated and
simplifies even more in the thermodynamic limit ($N, L \to \infty$ with
${N\over L}$ fixed) as well as in the long chain limit ($L \to \infty$
with ${N}$ fixed). The explicit form of \phider\ is
\eqn\first{
\Phi'_{ab}=-\delta_{ab}\left[{a'\over a}(\lambda_a) - 
{d'\over d}(\lambda_a) + \sum_{k\ne a}K(\lambda_a-\lambda_k) \right]
+K(\lambda_a-\lambda_b)
}
where
\eqn\eee{
K(\lambda)={f'\over f}(\lambda) + {f'\over f}(-\lambda)~~.
}
It turns out that $K$ is the kernel of the integral equation for the
density of rapidities in the thermodynamic limit. After using the
Bethe equations in the same limit, the coefficient of the Kronecker
$\delta$ becomes the density of rapidities. Combining everything,
$\Phi'$ is 
\eqn\aaa{
\Phi'_{ab}(\{\lambda\})=
L \left[\delta_{ab}\rho(\lambda_a)+{1 \over L}
K(\lambda_a-\lambda_b)\right]~~. 
}
The ratio $N/L$ enters here through the normalization of $\rho$.

In the long chain limit the result is even simpler because the
solutions of the Bethe equations are of the order of the length of the
chain, $\lambda_a = L/2\pi n_a$. Thus all terms depending on $K$ in
\first\ have subleading 
contributions and it is not necessary to solve the equation for
$\rho$. The matrix $\Phi'$ becomes diagonal, with entries given by the
first two terms in \first\ and its determinant is therefore 
\eqn\BMN{
\det \Phi'(\{\lambda\})= \prod_{a=1}^N \left[-{a'\over a}(\lambda_a) + 
{d'\over d}(\lambda_a)\right]~~.
}
Each term in the product is of order of the inverse length of the
chain. At first sight, this dependence as well as the extra factors in
\norm\ seem to be in contradiction with the known results concerning
long operators. Let us therefore briefly discuss the relation between
Bethe eigenstates and gauge theory operators.

Since we picked a normalized vacuum state $|0\rangle$, the
corresponding gauge theory operator must also be normalized. We
therefore have
\eqn\normvacuum{
|0\rangle~~~~~~~~\longleftrightarrow~~~~~~~~
{1\over \sqrt{L}}\Tr[(\phi^1)^L]~~.
}
If creation operators are present, the solution to the apparent
puzzle implied by \BMN\ is provided by the cyclicity of the trace and
the fact that the action of the creation operators $B(\lambda)$ brings
in additional rapidity-dependent factors. The relation between \eve\ 
and the non-normalized gauge theory operators is, roughly,
\eqn\rel{
|\Psi_N(\{\lambda\})\rangle=\prod_{i=1}^NB(\lambda_i)|0\rangle\sim 
\left[\prod_{i=1}^Nd(\lambda_i)\right]
\left[\prod_{i=1}^N{\gamma(\lambda_i)\over
\beta(\lambda_i)}\right]{{\cal O}^{{\rm non-norm}}\over \sqrt{L}}~~,
}
where ${\cal O}^{\rm non-norm}$ is, as usual, a linear combination of
traces of products of $(L-N)$ $\phi^1$ and $N$ $\phi^2$ fields while
$\sqrt{L}$ in the denominator comes from the normalization of
the vacuum state \normvacuum . 
Thus, $\left[\prod_{i=1}^Nd(\lambda_i)\right]$  cancels the similar
factor in the norm \norm\ of $|\Psi_N(\{\lambda\})\rangle$, while the
other factor contributes $L^{-N}$. Thus, combining \rel\ and \BMN\ we
find that the normalized BMN operators are related to the
non-normalized ones as
\eqn\eee{
{\cal O}^{\rm norm}\sim {1 \over L^{(N+1)/2}} {\cal O}^{\rm non-norm}~~.
}
Using the cyclicity of the trace we can always place a $\phi^2$ field 
as the first field inside the trace; this reduces the number of terms
in ${\cal O}^{\rm non-norm}$ by a factor of $L$ and simply multiplies the
resulting operator by $L$. We therefore find
\eqn\eee{
|\Psi_N(\{\lambda\})\rangle~~~\longleftrightarrow~~~
{\cal O}^{\rm norm}\sim { {\cal O}_0^{\rm non-norm}\over
L^{(N-1)/2}}~~~,~~ 
{\cal O}_0^{\rm non-norm}=\sum_p \Tr[\phi^2
K_p(\phi^1,\phi^2,\{\lambda\})] 
~~,
}
which is in agreement with the known gauge theory results. Here $K_p$
denotes some monomial in the fields with a $\lambda$-dependent
coefficient. It is
certainly possible to do a more detailed analysis and keep also the
subleading terms. 
We will see this explicitly in an example in the next section.

\newsec{Examples}

Let us illustrate the method of computing correlation functions with
some examples.
All examples will involve states created by two
creation operators, so we first normalize them. For a chain of 
length $L$ the corresponding 
solution of the Bethe equations \beq\ can be explicitly
found for the XXX chain 
\eqn\solrap{
\lambda_1=-\lambda_2=-{1\over 2}{\rm ctg}{\pi n\over L-1}~~,
}
where $\lambda_i$ are the shifted rapidities \shifted . 
Then equation \norm\ implies that the normalized state is
\eqn\optwoBMN{\eqalign{
&|\Psi_2^{\rm norm}(n)\rangle={\cal N}_2
B(\lambda_1)B(-\lambda_1)|0\rangle\cr
{\cal N}_2&={\lambda_1\over \sqrt{L(L-1)} }\left(\lambda_1^2+{1\over
4}\right)^{{1\over 2}-L}={4^{L-1}\over\sqrt{L(L-1)}}
\cos{\pi n\over L-1}\sin^{2(L-1)}{\pi n\over L-1}~~.
}}
Extracting the operators $B$ out of the monodromy matrix and acting
with them on the vacuum state leads to an
explicit expression for the (normalized) 
gauge theory operator associated to the
Bethe state  $|\Psi_2^{\rm norm}(n)\rangle$
\eqn\exploperator{
{\cal O}_n={1\over \sqrt{L-1}}\sum_{s=0}^{L-2}\,\cos{{\pi n(2s+1)\over
L-1}} \,\Tr[\phi^2(\phi^1)^s\phi^2(\phi^1)^{L-2-s}]~~,
}
in agreement with known results \BEOP . For other spin chains some
simple normalized gauge theory operators corresponding to Bethe states
have been constructed in \RR. Calculations similar to the
one leading to \exploperator\ yield (after solving the Bethe
equations) all finite length operators in the $SU(2)$ sector with 
definite one-loop anomalous dimensions.

\noindent{{\it Example One}}

A simple example of the type in which two operators have the same
length is provided by a three-point function of a generator of the 
Cartan subalgebra of the
$R$-symmetry group
\eqn\opcartan{
\Tr[\phi^1{\bar\phi}_1-\phi^2{\bar\phi}_2]
}
and two other operators. According to the discussion in \S 2, the 
tree-level coefficient of this correlation function is given by the
following expectation value: 
\eqn\eee{
\langle \Psi_N^{\rm norm}(\{\mu\})|\sum_{i=1}^L\sigma_i^z
|\Psi_N^{\rm norm}(\{\lambda\})\rangle=L\,\langle \Psi_N^{\rm
norm}(\{\mu\})| 
\sigma_1^z |\Psi_N^{\rm norm}(\{\lambda\})\rangle~~.
}
Using the solution \spin\ of the inverse scattering problem this
matrix element reduces to the matrix element of $A(0)-D(0)$ between 
the in- and out-states above. In this computation it is not 
necessary to use the regularization provided by the inhomogeneous chain,
because there are no potential singularities
from contractions.
This
matrix element was computed in \kitaninedet\ and the result is the
one expected on physical grounds
\eqn\eee{
\langle \Psi_N^{\rm norm}(\{\mu\})|\sum_{i=1}^L\sigma_i^z
|\Psi_N^{\rm norm}(\{\lambda\})\rangle =
(L-2N)\delta(\{\lambda\}-\{\mu\})~~, 
}
where $\delta(\{\lambda\}-\{\mu\})$ vanishes unless the set
$\{\lambda\}$ is the same as the set $\{\mu\}.$
Indeed, the operator \opcartan\ counts the difference between the
number of $\phi^1$ and $\phi^2$ fields. Since the action of $B$ replaces
$\phi^1$ with $\phi^2$, the result should be the difference between 
the number of $\phi^1$ fields in the vacuum and twice the number of
creation 
operators. Furthermore, since the Hamiltonian commutes with \opcartan ,
the latter does not mix the energy eigenstates, which explains the
delta function.

\noindent{{\it Example Two}}

A more complicated example is the extremal three-point function 
\eqn\extwo{
\langle \Tr[{{\bar \phi}_1}^{L-2}](x)  \Tr[{\bar \phi}_2^2](y)
{\cal O}_2(z)\rangle~~,
}
where ${\cal O}_2$ is the normalized operator corresponding to
$|\Psi^{\rm norm}_2(n)\rangle$. Using equation
\projshift\ it is not hard to find  that the tree level
normalization of \extwo\ is 
\eqn\second{
c_0=
\eqalign{
L
\sum_{i_1\dots i_{L-2}=1}^2~
{
{}_{{}_{L-2}}\langle 0 | \phi^{i_1}\dots
\phi^{i_{L-2}}\rangle
\langle{\bar\phi_2}{\bar\phi_2}
{\bar \phi}_{i_1}\dots {\bar
\phi}_{i_{L-2}}|\Psi^{\rm norm}_2(n)\rangle_{{}_L} 
}~~,
}
}
where the intermediate states are considered to be normalized\foot{In
fact, it is not hard to see from \S 2 that the intermediate states
are already unit normalized.}, the
indices on the vacuum states represent the length of the chain
describing them and the overall factor of $L$ comes from the sum over
the powers of the shift operator in \projshift. 
The scalar product
between the out vacuum state and the intermediate state forces the
intermediate state to also be the vacuum state $|0\rangle_{{}_{L-2}}$.
Thus, 
only one term in the sum survives
\eqn\eee{
c_0=
L~{\cal N}_2
~{
\langle{\bar\phi_2}{\bar\phi_2}
{\bar \phi}_{1}\dots {\bar
\phi}_{1}|B(\lambda_1)B(-\lambda_1)|0\rangle_{{}_L}~~,
}
}
where ${\cal N}_2$ is the normalization constant of the in-state
\optwoBMN. Using the solution \spin\ of the inverse scattering problem
and the fact that \solrap\ represents shifted rapidities,
the remaining scalar product reduces to\foot{As discussed before, the
arguments of the $C$ operators above must be 
the values of the rapidities for which the $R$-matrix becomes the
permutation operator. For the shifted rapidities this is
$\lambda_0={i\over 2}$.}
\eqn\eee{
c_0=L~{\cal N}_2
~\langle0|\, 
C({{\textstyle{i\over 2}}})C({{\textstyle{i\over 2}}})\,
B(\lambda_1)B(-\lambda_1)|0\rangle~~.
}
Using \adc\ and the regularization provided by the anisotropic chain
or the general expression for the scalar product between an energy
eigenstate and an arbitrary state \sp, we find
\eqn\aaa{\eqalign{
&\langle 0|C({{\textstyle{i\over 2}}}) C({{\textstyle{i\over 2}}}) B(\lambda_1) B(-\lambda_1)
|0\rangle
=2\left(\lambda^2_1+{1\over 4}\right)^{L-1}
={ 2^{3-2L}\sin^{2(1-L)}{\pi n\over L-1}}
}~~.
}
Combining this with the normalization of the in-state \optwoBMN\ 
leads to the three-point function coefficient
\eqn\cnot{
c_0=\sqrt{L\over L-1} {4\lambda_1\over \sqrt{1+4\lambda_1^2}}
=2\sqrt{L\over L-1} \cos {\pi n\over L-1}~~.
}
Similar coefficients have been computed \GMIT\
for some operators dual to
string modes in plane wave background. 
The equation \cnot\ is more general since it is valid 
for operators of arbitrary length.

As a side remark, it is not hard to see using the analysis of \S 2
that the same tree-level three-point function coefficient arises, up
to some numerical coefficient, if we
replace $\Tr[{\bar\phi}_2^2]$ with $\Tr[(\phi^1)^k{\bar\phi}_2^2]$ for
any $k\le L-4$ and also appropriately adjust the length of the chain
describing the out-state. 

\fig{Various diagrams contributing to \extwo\ at one-loop level.}
{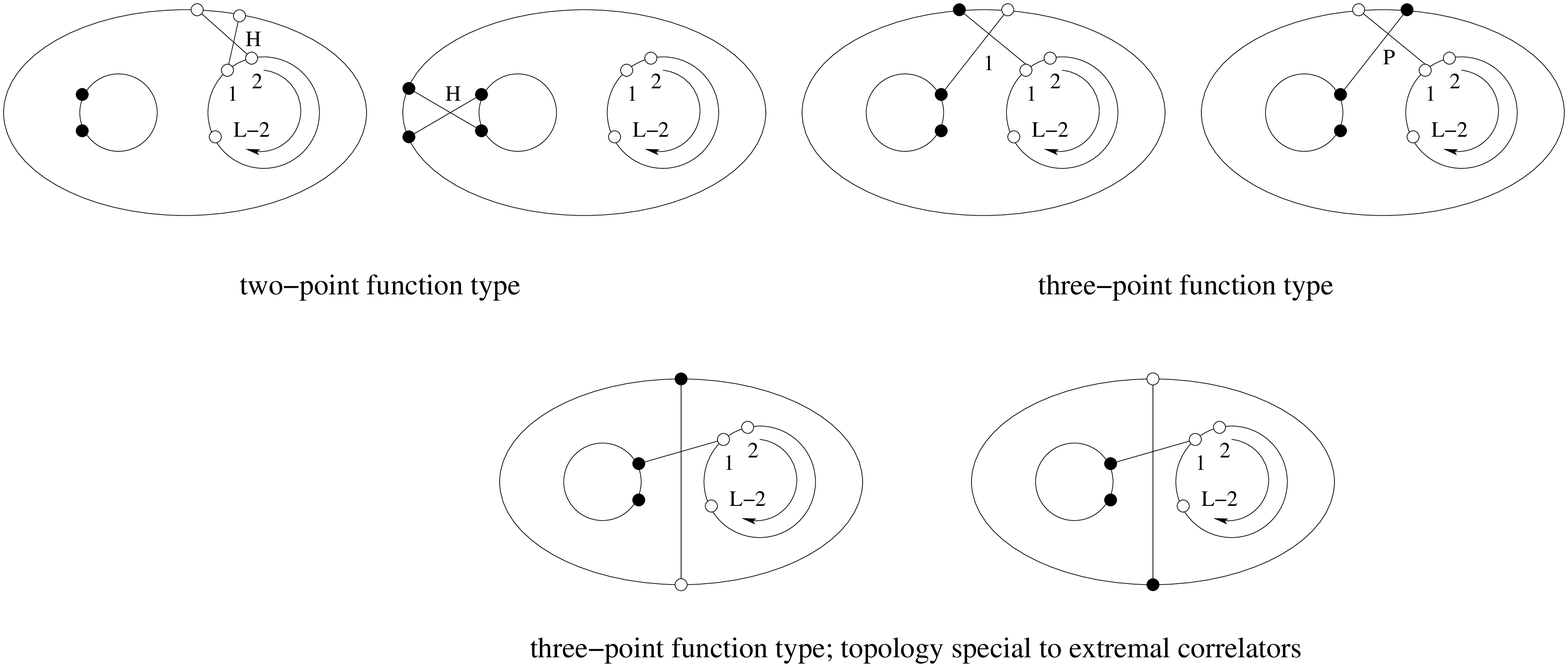}{5.2in}

Using the analog of the equations \twosix\ and \threesix\ for the
$SU(2)$ sector it is not hard to find the coefficients of the one-loop
corrections to \extwo. The relevant gauge theory Feynman diagrams,
grouped as discussed in \S 2, are listed in figure 3.
The white dots correspond to $\phi_1$ while black
dots correspond to $\phi_2$.
Some of the contributions are particularly simple. Since both 
$\Tr[{\bar\phi}_1^{L-2}]$ and $\Tr[{\bar\phi}_2^2]$ are BPS operators,
they are annihilated by the Hamiltonian of the spin chain and the
contribution of the analog of equation \twosix\ (first two diagrams in
figure 3) vanishes. The contribution of the analog of \threesix\  does
not vanish and it is a sum of two terms. The simplest contribution
(third diagram in figure 3) comes from the identity operator in the
Hamiltonian and contributes $c_0$ to the coefficient of
$\ln{|z-x|^2|z-y|^2/|x-y|^2}$. A similar contribution comes from 
the interactions typical to extremal three-point functions (fifth and
sixth diagrams above).

The final contribution (fourth diagram in figure 3) comes from the
permutation operator and  is proportional to  
\eqn\last{
C_P^{(1)}=
\langle 0|C(\textstyle{i\over 2})(A(\textstyle{i\over 2})+
D(\textstyle{i\over 2}))C(\textstyle{i\over 2})
|\Psi^{\rm norm}_2(n)\rangle~~.
}
This matrix element is a special case of a spin-spin correlation
function and measures the probability of having two down spins with an
up spin between them. The computation of this scalar product is
relatively easy. As before, we need to regularize it by shifting the
arguments of the operators in \last\ at arbitrary
positions. It is not hard to see that the nontrivial normalization
factor of the regularized state is irrelevant, so we will not compute
it explicitly.
Defining\foot{Clearly, $S({i\over 2})=c_0$.}
\eqn\eee{\eqalign{
S(\xi)&=\langle 0|C(\xi)C({\textstyle{i\over 2}})
|\Psi^{\rm norm}_2(n)\rangle\cr
&={\lambda_1\sqrt{1+4\lambda_1^2}\left({1\over
2}-i\xi\right) \over
\sqrt{L(L-1)} (\lambda_1^2-\xi^2)}
\left[\left({1\over 2}-i\xi\right)^{L-1}
-\left(-{1\over 2}-i\xi\right)^{L-1}\right]
}}
which can be easily computed using \sp , the 
scalar product \last\ is given by
\eqn\exp{
C_P^{(1)}=\left(a({\textstyle{i\over 2}})+d({\textstyle{i\over 2}})
\right)
S({\textstyle{i\over 2}})+ i
\left[
a(\xi)^2\partial_\xi \left({S(\xi)\over a(\xi)}\right)-
d(\xi)^2 \partial_\xi \left({S(\xi)\over d(\xi)}\right) 
\right]
\Bigg|_{\xi={i\over 2}} ~~.
}
Using the explicit form of $a$, $d$ and $S(\xi)$ we find that the last
one-loop three-point function coefficient is given by 
\eqn\eee{
C_P^{(1)}=c_0\left[1-4
\sin^2{\pi n\over L-1}\right]=\sqrt{L\over L-1}\cos{3\pi n\over L-1}~~.
}
Collecting the contributions of all diagrams in figure 3 yields the 
one-loop correction to the three-point function \extwo.

\bigbreak\bigskip\bigskip
\bigbreak\bigskip\bigskip
\centerline{\bf Acknowledgments}\nobreak

We have benefited from helpful discussions with David Berenstein, Joe
Polchinski, Marcus Spradlin, Mark Srednicki and Li-Sheng Tseng.
This work was supported in part by the National Science Foundation
 under Grants PHY99-07949 
(AV) and PHY00-98395 (RR), as well as by the DOE under Grant
No.~91ER40618 (RR).  Any opinions, findings, and conclusions or
recommendations expressed in this material are those of the authors
and do not necessarily reflect the views of the National Science
Foundation.

\listrefs

\end